\documentclass[prb,aps,superscriptaddress,showpacs,floatfix]{revtex4}

\usepackage{amsmath,amssymb}
\usepackage{graphicx}
\usepackage{version}
\DeclareGraphicsExtensions{.eps,.ps}

\newcommand{\bQ}{{\bf Q}}
\newcommand{\bk}{{\bf k}}
\newcommand{\bK}{{\bf K}}

\newcommand{\bp}{{\bf p}}
\newcommand{\bq}{{\bf q}}

\newcommand{\ba}{{\bf a}}

\newcommand{\bb}{{\bf b}}

\newcommand{\beqa}{\begin{eqnarray}}
\newcommand{\eeqa}{\end{eqnarray}}

\begin{document}

\title
{Influence of Landau-level mixing on Wigner crystallization in graphene}
\author{C.-H. Zhang}
\affiliation{Department of Physics, Indiana University-Purdue
University Indianapolis, Indianapolis, Indiana 46202, USA}
\author{Yogesh N. Joglekar}
\affiliation{Department of Physics, Indiana University-Purdue
University Indianapolis, Indianapolis, Indiana 46202, USA}
\date{\today}
\begin{abstract}
Graphene, with its massless linearly-dispersing carriers, in the 
quantum Hall regime provides an instructive comparison with conventional 
two-dimensional (2D) systems in which carriers have a nonzero band mass and 
quadratic dispersion. 
We investigate the influence of Landau level mixing in graphene on Wigner 
crystal states in the $n^\mathrm{th}$ Landau level obtained using single 
Landau level approximation. We 
show that the Landau level mixing does not qualitatively change the phase 
diagram as a function of partial filling factor $\nu$ 
in the $n^\mathrm{th}$ level. We find that the 
inter-Landau level mixing, quantified by relative occupations of the two 
Landau levels, $\rho_{n+1}/\rho_{n}$, oscillates around $2\%$ and, 
in general, remains small ($< 4\%$) irrespective of the Landau level 
index $n$. Our results show that the single Landau level 
approximation is applicable in high Landau levels, even though the energy 
gap between the adjacent Landau levels vanishes. 
\end{abstract}

\maketitle

\section{Introduction}

Wigner crystallization, where the density profile of carriers in a system 
develops a periodic spatial modulation spontaneously, is a classic example 
of interplay between (classical) repulsive potential energy and the (quantum) 
kinetic energy associated with localization of carriers as the density of 
carriers is varied.~\cite{wigner1934,fetter,tanatar1989} 
Although predicted in 1934~\cite{wigner1934} this phenomenon has defied 
direct experimental observation in bulk systems and 
conventional 2D systems. In quantum Hall systems, where the kinetic 
energy of carriers is 
quantized and quenched, Wigner crystallization is induced by a competition 
between the electrostatic and exchange interactions as the partial filling 
factor $\nu$ in a given Landau level is varied. (In the quantum Hall regime, 
Wigner crystallization depends only on the filling factor and can occur at 
any carrier density.~\cite{Das97}) Wigner crystallization in the lowest 
Landau level has been 
inferred via transport measurements,~\cite{willett1988} and the anisotropic 
transport observed~\cite{mikestripes} in high Landau levels can be 
interpreted~\cite{stripetheory} in terms of anisotropic Wigner crystal 
ground states. A direct observation of the Wigner crystal, via 
local carrier density, however, has not yet been possible. 
Graphene, with its massless carriers on the surface, is a 
unique and ideal candidate for this purpose.~\cite{Novoselov05,yacobi} 
Recent studies, using Hartree-Fock mean-field theory in the 
single-Landau-level approximation (SLLA)~\cite{Zhang07} or 
exact diagonalization in the single-Landau-level subspace~\cite{Wang07} 
have predicted 
that Wigner crystal states will appear as ground states over a range of 
partial filling factor $\nu$ in a given Landau level. In this paper, we 
examine the validity of the single-Landau-level approximation. 

Let us first recall the relevant results for a conventional 2D system in 
perpendicular magnetic field $B$ with partial filling factor $\nu\leq 1$ 
in the Landau level $n$. Thus, the actual filling factor 
for spinless carriers (with no other degeneracies) is $n+\nu$. For this 
system, the difference between energies of the adjacent Landau levels is 
$\Delta E_n=E_{n+1}-E_{n}=\hbar\omega_c$ where $\omega_c=eB/mc$ is the 
cyclotron frequency, $m\sim 0.5m_e-0.1 m_e$ is the band mass of the 
carriers, and $m_e$ is the bare electron mass. We remind the Reader that 
$\Delta E_n=\hbar^2/ml_B^2$ is (approximately) the 
quantum kinetic energy of a particle with mass $m$ in a box with size 
$l_B=\sqrt{\hbar c/eB}$. The Coulomb interaction that causes transitions 
between different Landau levels has a typical energy scale $V_c=e^2/\epsilon 
l_B$ where $\epsilon\sim 10$ is the dielectric constant. Therefore, the 
ratio of these two energy scales, $V_c/\Delta E_n=l_B/a_B$ where 
$a_B=\hbar^2\epsilon/me^2$ is the Bohr radius of the carriers in the 
material. Since $a_B$ is independent of the magnetic 
field and the magnetic length $l_B\propto 1/\sqrt{B}$, as 
$B\rightarrow\infty$ the amplitude for inter-Landau level transitions 
vanishes and the SLLA becomes a good 
approximation.~\cite{MacDonald84} A corresponding analysis for graphene 
shows the stark difference between 
the two systems. The gap between the adjacent Landau level energies 
in graphene is $\Delta E_n=E_{n+1}-E_n=\hbar\omega[\sqrt{2(n+1)}-\sqrt{2n}]$ 
where $\omega=v_G/l_B$ is the cyclotron frequency, $v_G\sim c/300$ is 
the speed of massless carriers in graphene, and $c$ is the speed of light. 
It follows that the ratio 
\begin{equation}
\label{eq:g_n}
g_n=\frac{V_c}{\Delta E_n}=\frac{e^2}{\epsilon\hbar v_G}\frac{1}
{\sqrt{2(n+1)}-\sqrt{2n}}
\end{equation}
is independent of the magnetic field and diverges, $g_n\sim\sqrt{2n}$, 
as $n\rightarrow\infty$. Therefore, inter-Landau level transitions become 
increasingly important as the Landau level index $n$ increases, 
irrespective of the magnetic field; even in the lowest Landau level, the 
ratio $g\sim e^2/\epsilon\hbar v_G=\alpha_G\sim 1$ is not 
small ($\alpha_G$ is the fine structure constant for graphene). This 
analysis suggests that the SLLA is not reliable in graphene for any $B$ and 
that it gets worse with increasing $n$ since the energy gap 
$\Delta E_n\rightarrow 0$. In the following we show that, contrary to the 
expectations from a simple analysis presented above, the effect of 
Landau level mixing in graphene remains small and SLLA remains applicable. 

The outline of the paper is as follows. In Sec.~\ref{sec:HFA}, we briefly 
describe the Hartree-Fock approximation with Landau-level mixing and outline 
our approach. The details presented in this section are essentially 
identical to those in our earlier work.~\cite{Zhang07} 
In Sec.\ \ref{sec:numerics}, we present the results obtained without and 
with Landau-level mixing. We find that the Landau-level mixing does not 
qualitatively change the phase diagram of the system. We quantify the 
mixing using 
off-diagonal self-energy and relative occupation of Landau levels $n$ 
and $n+1$. We compare the results for Landau level mixing as a function of $n$ 
in graphene with those for conventional 2D systems. We summarize our 
conclusions in Sec.~\ref{sec:disc}. 


\section{Microscopic Hamiltonian and Hartree-Fock Approximation}
\label{sec:HFA}

Let us consider graphene in a strong perpendicular magnetic field $B$ in the 
quantum Hall regime. The single-particle states of the non-interacting 
system are given by $|n,k,\sigma\rangle$ where $(n,k)$ denote the Landau 
level and intra-Landau level indices, and $\sigma=\pm$ correspond to the two 
inequivalent valleys, $\bK$ and $\bK'=-\bK$, in the Brillouin zone. The 
details presented in this section follow closely Ref.[\onlinecite{Zhang07}]. 
The Hamiltonian for the system, including the Coulomb interaction is 
\begin{align}
\label{eq:H_int} 
\hat{H} & =N_{\phi}\sum_{n\sigma}(E_n-\mu)\hat{\rho}^{\sigma,\sigma}_{n,n}(0) 
+ \frac{1}{2A}\sum_{\sigma_i n_j\bq} V(\bq)
\mathcal{F}_{n_1,n_2}(-\bq)\hat{\rho}^{\sigma_1,\sigma_1}_{n_1,n_2}(\bq)
\mathcal{F}_{n_3,n_4}(-\bq)\hat{\rho}^{\sigma_3,\sigma_3}_{n_3,n_4}(\bq) 
\end{align}
where $A$ is the area of the sample, $\mu$ is the chemical potential, 
$V(\bq)=2\pi e^2/\epsilon q$ is the Coulomb interaction in graphene 
($\epsilon\sim 2-5$), and
\begin{align}
\label{eq:rho}
\hat{\rho}^{\sigma,\sigma'}_{n,n'}(\bq)=\frac{1}{N_\phi}\sum_{k,k'}
e^{-\frac{i}{2}q_x(k+k')l_B^2}\delta_{k,k'-q_y} c^{\dagger}_{nk\sigma}
c_{n'k'\sigma'}.
\end{align}
with $c^{\dagger}_{nk\sigma} (c_{nk\sigma})$ representing the creation 
(annihilation) operator for state $|n,k,\sigma\rangle$. Eq.(\ref{eq:rho}) 
is related to the density matrix operator in the momentum space 
\begin{align}
\label{eq:rhototal}
\hat{\rho}(\bq)=\sum_{nn'\sigma\sigma'}
{\cal F}_{n,n'}(-\bq)\hat{\rho}_{n,n'}^{\sigma,\sigma'}(\bq),
\end{align}
where the form factor for graphene (with $n,n'\geq 0$) is given 
by~\cite{Zhang07} 
\begin{align}
\label{eq:ffgraphene}
{\cal F}_{n,n'}(\bq)&=\delta_{n,0}\delta_{n',0}
F_{0,0}(\bq)+\frac{1}{\sqrt{2}}\delta_{nn',0}(1-\delta_{n+n',0})
F_{n,n'}(\bq)
\nonumber\\
&+\frac12(1-\delta_{nn',0})\left[F_{n,n'}(\bq)+F_{n-1,n'-1}(\bq)\right].
\end{align}
We recall that $\mathcal{F}_{n,n'}(\bq)$ is a linear combination of the form 
factors for a conventional 2D system,~\cite{Zhang07}
\begin{align}
F_{n\geq n'}(\bq)=\sqrt{\frac{n'!}{n!}}\left[\frac{(iq_x-q_y)}
{\sqrt{2}}\right]^{(n-n')}L_{n}^{(n-n')}\left(\frac{q^2}{2}\right)e^{-q^2/4}
\end{align}
where $L_{n}^m(x)$ is the generalized Laguerre polynomial and 
$F_{n\leq n'}(\bq)=F^{*}_{n',n}(-\bq)$.

The derivation of the mean-field Hamiltonian using Hartree-Fock approximation 
is straightforward~\cite{MacDonald84} and gives
\begin{align}
\label{eq:HF}
\hat{H}_{HF}&=N_{\phi}\sum_{\sigma n}(E_{n}-\mu)\hat{\rho}^{\sigma,\sigma}
_{n,n}(0)
+N_{\phi}\sum_{\sigma_i n_j \bq}U_{\sigma_1n_1,\sigma_2
n_2}(\bq)\hat{\rho}^{\sigma_1,\sigma_2}_{n_1,n_2}(\bq)
\end{align}
where $U_{\sigma_1n_1,\sigma_2 n_2}(\bq)=H_{\sigma_1n_1,\sigma_2
n_2}(\bq)-X_{\sigma_1n_1,\sigma_2n_2}(\bq)$. The self-consistent 
electrostatic and exchange potentials are given by 
\begin{align}
H_{\sigma_1n_1,\sigma_2n_2}(\bq)
&=\delta_{\sigma_1,\sigma_2}\sum_{n_3n_4\sigma}H_{n_1n_3,n_2n_4}(\bq)
\rho^{\sigma,\sigma}_{n_3,n_4}(-\bq),\\
X_{\sigma_1n_1, \sigma_2n_2}(\bq)
&=\sum_{n_3n_4}X_{n_1n_3,n_2n_4}(\bq)\left[\delta_{\sigma_1,\sigma_2}
\rho^{\sigma_1,\sigma_1}_{n_3,n_4}(-\bq)+\delta_{\sigma_1,\bar{\sigma}_2}
\rho^{\bar{\sigma_1},\sigma_1}_{n_3,n_4}(-\bq)\right],
\end{align}
where
\begin{align}
\label{eq:hartree}
H_{n_1n_3,n_2n_4}(\bq)&= \frac{1}{2\pi
l_B^2}V(\bq)(1-\delta_{\bq,0})
\mathcal{F}_{n_1,n_2}(-\bq)\mathcal{F}_{n_3,n_4}(\bq),\\
\label{eq:fock}
X_{n_1n_3,n_2n_4}(\bq) &=\int\frac{{d\bk}}{(2\pi)^2}V(\bk)
e^{-il_B^2\bk\times\bq\cdot\hat{z}}
\mathcal{F}_{n_1,n_4}(\bk)\mathcal{F}_{n_3,n_2}(-\bk),
\end{align}
and $\rho^{\sigma_1,\sigma_2}_{n_1,n_2}(\bq)=\langle\hat{\rho}
^{\sigma_1,\sigma_2}_{n_1,n_2}(\bq)\rangle$ are the density matrix elements 
which should be determined self-consistently from Eq.(\ref{eq:HF}). The 
density matrix is obtained from the 
equal-time limit ($\tau\rightarrow 0^{-}$) of the single-particle Green's 
function
\begin{align}
G^{\sigma_1,\sigma_2}_{n_1,n_2}(k_1,k_2;\tau)=-\langle\mbox{T}
c_{n_1k_1\sigma_1}(\tau)c^\dagger_{n_2k_2\sigma_2}(0)\rangle.
\end{align}
The equation of motion for the Green's function in Fourier space 
is given by~\cite{Zhang07}
\begin{align}
\label{eq:gf}
\delta_{\sigma_1,\sigma_{2}}\delta_{n_1,n_2}\delta_{\bq,0}
&=\left[i\omega_n-\left(E_{n_1}-\mu\right)\right]
G^{\sigma_1,\sigma_2}_{n_1,n_2}(\bq,i\omega_n)-\sum_{\sigma_3n_3\bq^\prime}
\Sigma_{\sigma_1n_1,\sigma_3n_3}(\bq,\bq^\prime)G^{\sigma_3,\sigma_2}_
{n_3,n_2}(\bq^{\prime},i\omega_n) 
\end{align}
and the Hartree-Fock self-energy matrix is ($\bp=\bq-\bq'$) 
\begin{align}
\label{eq:sigma}
\Sigma_{\sigma_1n_1,\sigma_3n_3}(\bq,\bq^\prime)&=\sum_{m_1m_3}
\left\{\left[ 
H_{n_1m_1,n_3m_3}(-\bp)\rho_{m_3,m_1}(\bp)-X_{n_1m_1,n_3m_3}(-\bp)
\rho^{\sigma_1,\sigma_1}_{m_3,m_1}(\bp)\right]\delta_{\sigma_1,\sigma_3}
\right.\nonumber\\
&\quad\left. -X_{n_1m_1,n_3m_3}(-\bp)\rho^{\bar{\sigma}_1,\sigma_1}
_{m_3,m_1}(\bp)\delta_{\sigma_3,\bar{\sigma}_1}\right\}
e^{\frac{i}{2}l_B^2\bq\times\bq^\prime\cdot\hat{z}},
\end{align}
where we have defined $\rho_{m_3,m_1}(\bp)=\Sigma_{\sigma}
\rho^{\sigma,\sigma}_{m_3,m_1}(\bp)$. 

In single Landau level approximation for the $n^\mathrm{th}$ level, all 
Landau level indices in Eq.(\ref{eq:gf}) are the same, $n_1=n_2=n_3=n$. To 
account for the inter-Landau level transitions, we restrict the indices 
to $n$ and $n+1$. The Green's function in the Landau-level space then becomes 
a 2$\times$2 matrix,
\begin{align}
\tilde{G}^{\sigma_1,\sigma_2}(\bq,i\omega_n)=\left[\begin{array}{cc}
G^{\sigma_1,\sigma_2}_{n,n} & G^{\sigma_1,\sigma_2}_{n,n+1}\\
G^{\sigma_1,\sigma_2}_{n+1,n} & G^{\sigma_1,\sigma_2}_{n+1,n+1}\\
\end{array}\right](\bq,i\omega_n) 
\end{align}
and similarly the self-energy matrix 
$\tilde{\Sigma}_{\sigma_1,\sigma_2}(\bq,\bq')$ is a 2$\times$2 matrix in 
the Landau-level space. The equation of motion for the Green's function, 
Eq.(\ref{eq:gf}), becomes 
\begin{align}
\label{eq:llmatrix}
\delta_{\sigma_1,\sigma_2}\delta_{\bq,0}=\left[i\omega+\mu\right]
\tilde{G}^{\sigma_1,\sigma_2}(\bq,i\omega_n)-\sum_{\sigma_3\bq^\prime}\left[
\tilde{\Sigma}_{\sigma_1,\sigma_3}(\bq,\bq^\prime)+
\tilde{E}\delta_{\bq,\bq'}\delta_{\sigma_1,\sigma_3}\right]
\tilde{G}^{\sigma_3,\sigma_2}(\bq^\prime,i\omega_n)
\end{align}
where the kinetic energy matrix in the Landau level space is diagonal, 
$\tilde{E}=\mathrm{diag}(E_n,E_{n+1})$. We solve Eq.(\ref{eq:llmatrix}) by 
obtaining the eigenvalues and eigenvectors 
\begin{align}
\label{eq:eigen} \sum_{\sigma_3\bq^\prime}\left[
\tilde{\Sigma}_{\sigma_1,\sigma_3}(\bq,\bq^\prime)+\tilde{E}
\delta_{\bq,\bq'}\delta_{\sigma_1,\sigma_3}\right]
\tilde{V}_{\sigma_3}(\bq^\prime,k)=\omega_k
\tilde{V}_{\sigma_1}(\bq,k).
\end{align}
Here $\tilde{V}^{\dagger}_{\sigma}(\bq,k)=
\left[V^{*}_{\sigma,n}(\bq,k),V^{*}_{\sigma,n+1}(\bq,k)\right]$ is the 
eigenvector with eigenvalue $\omega_k$. We can construct the self-consistent 
mean-field Green's function using these eigenvectors~\cite{Zhang07}
\begin{align}
\tilde{G}^{\sigma_1,\sigma_2}(\bq,i\omega_n)=\sum_k
\frac{\tilde{V}_{\sigma_1}(\bq,k)\tilde{V}_{\sigma_2}^\dagger(0,k)}
{i\omega_n-\omega_k+\mu}
\end{align} 
which, in turn, leads to the self-consistent density matrix
\begin{align}
\label{eq:density}
\rho^{\sigma_1,\sigma_2}_{n_1,n_2}(\bq)=\sum_kV_{\sigma_2,n_2}(\bq,k)
V^*_{\sigma_1,n_1}(0,k)f(\omega_k-\mu),
\end{align}
where $f(x)=\theta(-x)$ denotes the Fermi function at zero temperature. 
The chemical 
potential $\mu$ is determined by the constraint that the total occupation 
in the two Landau levels is equal to the partial filling factor,  
\begin{align}
\sum_{\sigma}[\rho^{\sigma,\sigma}_{n,n}(0)+\rho^{\sigma,\sigma}_{n+1,n+1}
(0)]=\nu.
\end{align}
Using the self-consistent density matrix (\ref{eq:density}), we calculate 
the Hartree-Fock mean-field energy $E_{HF}$ for various trial lattice 
configurations to obtain the ground state crystal structure. 


\section{Results}
\label{sec:numerics}

We consider mean-field Wigner crystal lattices with two primitive lattice 
vectors $\ba_1=(a,b/2), \ba_2=(0,b)$ and define the lattice anisotropy as 
$\gamma=b/a$. Note that the triangular lattice ($\gamma=2/\sqrt{3}=1.15$) 
and quasi-striped states ($\gamma\rightarrow 0$) are special cases of the 
general anisotropic lattice defined by these primitive 
vectors. The 
lattice constants $a$ and $b$ are determined by the constraint that a unit 
cell contains $N_e$ electrons, and are given by 
$a=l_B\sqrt{2\pi N_{e}/\nu\gamma}$ and $b=a\gamma$. The reciprocal lattice 
vectors are $\bQ_{mn}=m\bb_1+n\bb_2$ where $\bb_1=(2\pi/a)(1,0)$ 
and $\bb_2=(2\pi/a)(-1/2,1/\gamma)$ are the reciprocal lattice basis vectors.  
We determine the optimal lattice structure by choosing the 
$\gamma$ ($0<\gamma\le2/\sqrt{3}$) and $N_e$ that minimize the mean-field 
energy $E_{HF}$. 
In the following, we denote crystals with one electron per unit 
cell, $N_e=1$, as Wigner crystals and those with $N_e\ge 2$ per unit cell as 
bubble crystals.~\cite{Koulakov96,goerbig04} We first calculate the 
self-consistent density matrix without Landau level mixing, 
$\rho^{\sigma_1,\sigma_2}_{n,n}(\bq)\ne0$ and 
$\rho^{\sigma_1,\sigma_2}_{n+1,n}(\bq)=0
=\rho^{\sigma_1,\sigma_2}_{n+1,n+1}(\bq)$. We then use that matrix as the 
initial point for the density matrix with Landau-level mixing. 

\begin{figure}[thb]
\begin{center}
\begin{minipage}{20cm}
\hspace{-15mm}
\begin{minipage}{9cm}
\includegraphics[width=8cm]{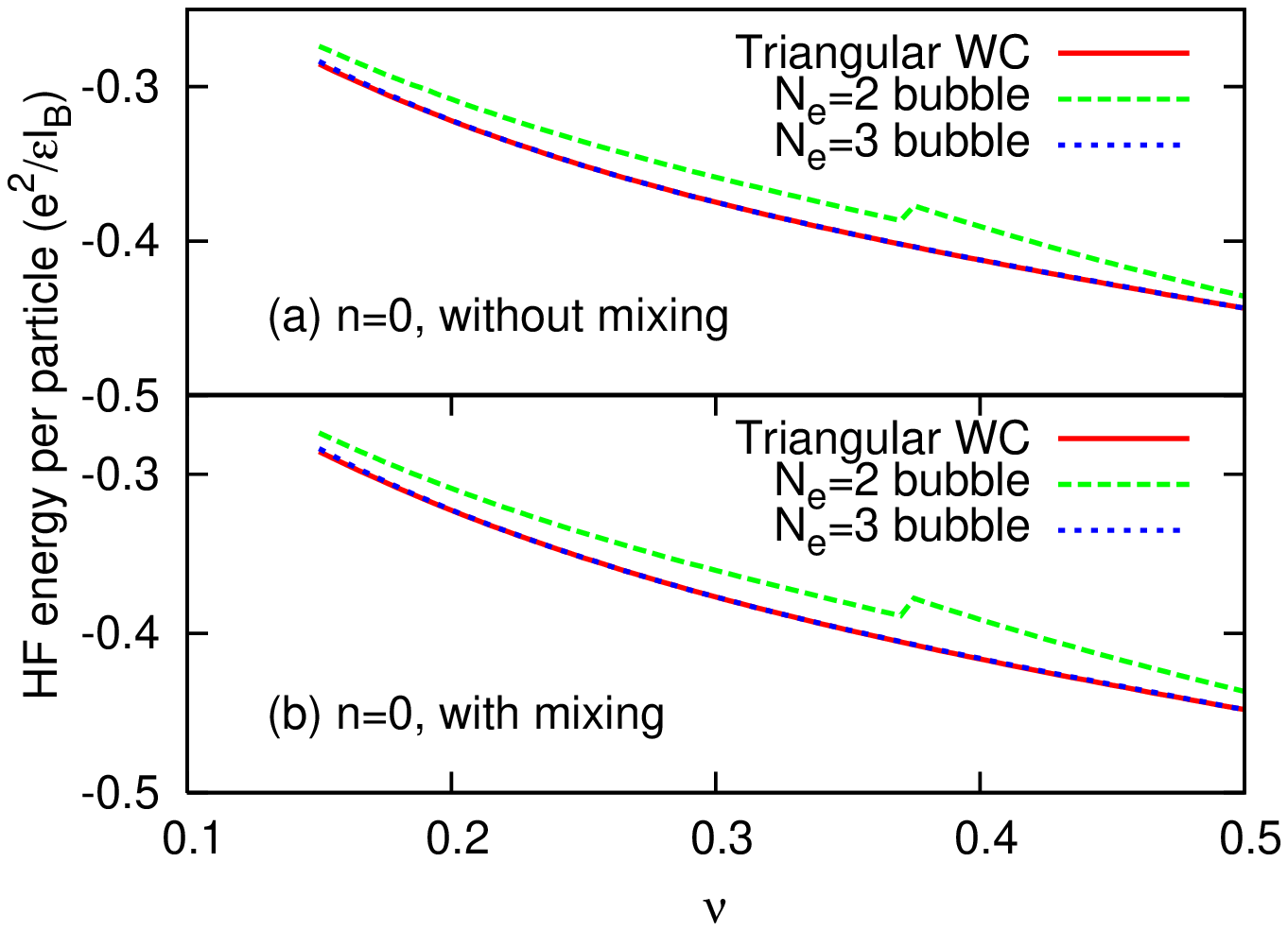}
\end{minipage}
\begin{minipage}{9cm}
\includegraphics[width=8cm]{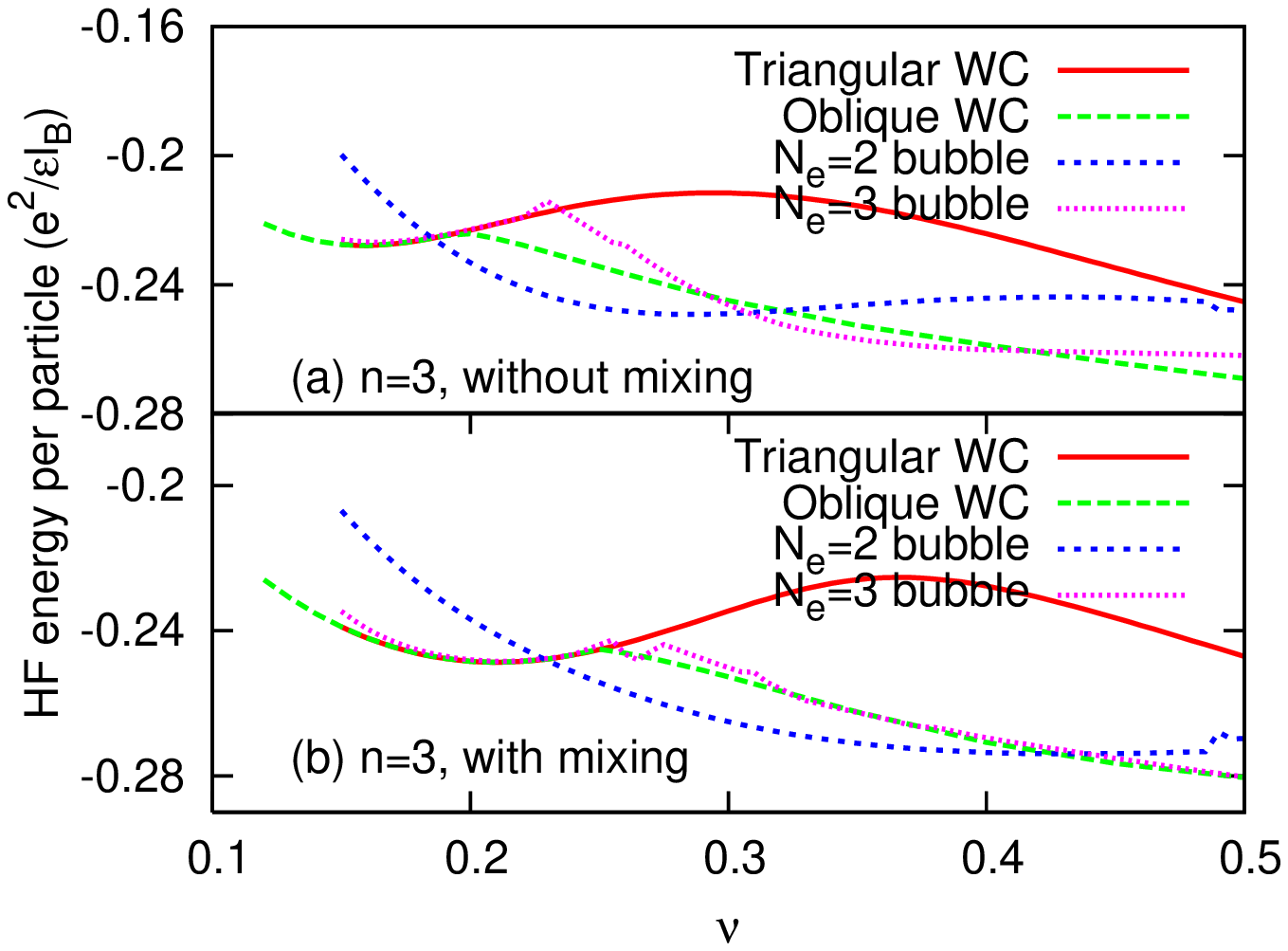}
\end{minipage}
\end{minipage}
\caption{(Color Online) Ground state energy per particle, measured in 
units of $e^2/\epsilon l_B$, for different crystal structures in the 
$n=0$ (left) and $n=3$ (right) Landau levels in graphene. The top panel (a) 
shows results without Landau level mixing, whereas the bottom panel (b) 
shows results with mixing. We see that, in each case, the 
{\it ground state energy} is lowered due to Landau level 
mixing.~\cite{Sakurai} Note that the phase-diagram is qualitatively 
unchanged.} 
\label{fig:mixing}
\end{center}
\vspace{-5mm}
\end{figure}
Figure~\ref{fig:mixing} shows mean-field energy per particle for various 
lattice structures as a function of $\nu$ for Landau level $n=0$ (left) 
and $n=3$ (right). We note that the {\it ground state energy} for a 
given $\nu$ is lowered by the Landau level mixing, as expected from 
perturbation theory.~\cite{Sakurai} For $n=0$, we find that a triangular 
Wigner crystal is the mean-field ground state with or without Landau 
level mixing. For $n=3$, we find that the triangular 
lattice remains a ground state for higher values of $\nu$ when the 
inter-Landau level mixing is taken into account. Overall, the phase diagram 
of the system remains qualitatively unchanged. 

\begin{figure}[thb]
\includegraphics[width=10cm]{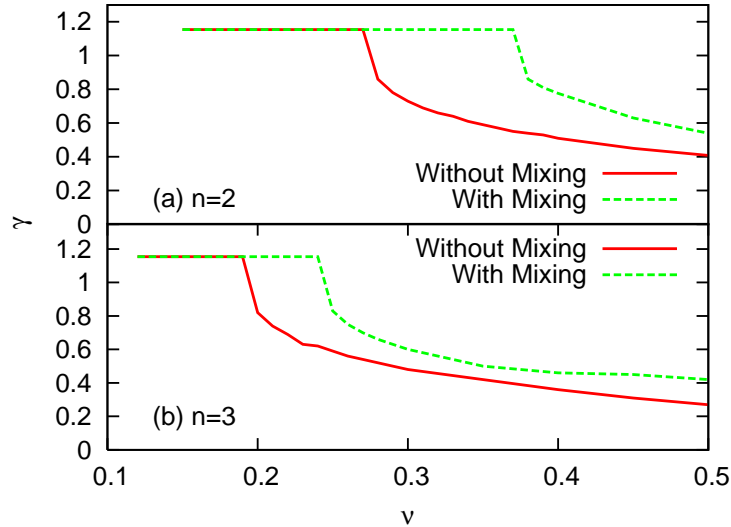}
\caption{(Color Online) Ground state lattice anisotropy $\gamma(\nu)$ in 
graphene for $n=2$ (top) and $n=3$ (bottom) Landau levels. The solid (red) 
line shows the result without mixing and the dashed (green) line 
indicates the result with Landau level mixing. $\gamma=2/\sqrt{3}=1.15$ 
corresponds to a triangular lattice whereas $\gamma\leq 0.5$ corresponds 
to highly anisotropic Wigner crystal or quasi-striped states.} 
\label{fig:gamma}
\end{figure}
The most visible effect of inter-Landau level mixing is the systematic 
up-shift of critical values of $\nu$ at which transitions from one crystal 
structure to another occur. For example, at $n=3$ 
the transition from an isotropic Wigner crystal to an $N_e=2$ anisotropic 
bubble state occurs at $\nu\sim 0.20$ without Landau-level mixing; this 
critical value is shifted upwards to $\nu\sim 0.25$ when the mixing is 
taken into account (Figure~\ref{fig:mixing}). This shift is also visible
in the lattice anisotropy $\gamma(\nu)$ for the ground state crystal 
structure, shown in Fig.\ \ref{fig:gamma}. At small $\nu$, the lattice is 
triangular and $\gamma=2/\sqrt{3}=1.15$ is a constant. At higher values 
of $\nu$, the anisotropy increases leading 
to a quasi-striped structure for the ground state. We see from 
Fig.~\ref{fig:gamma} that the region of stability of the triangular lattice 
increases when inter-Landau level transitions are taken into account. 

\begin{figure}[thb]
\begin{minipage}{20cm}
\hspace{-15mm}
\begin{minipage}{9cm}
\includegraphics[width=9cm]{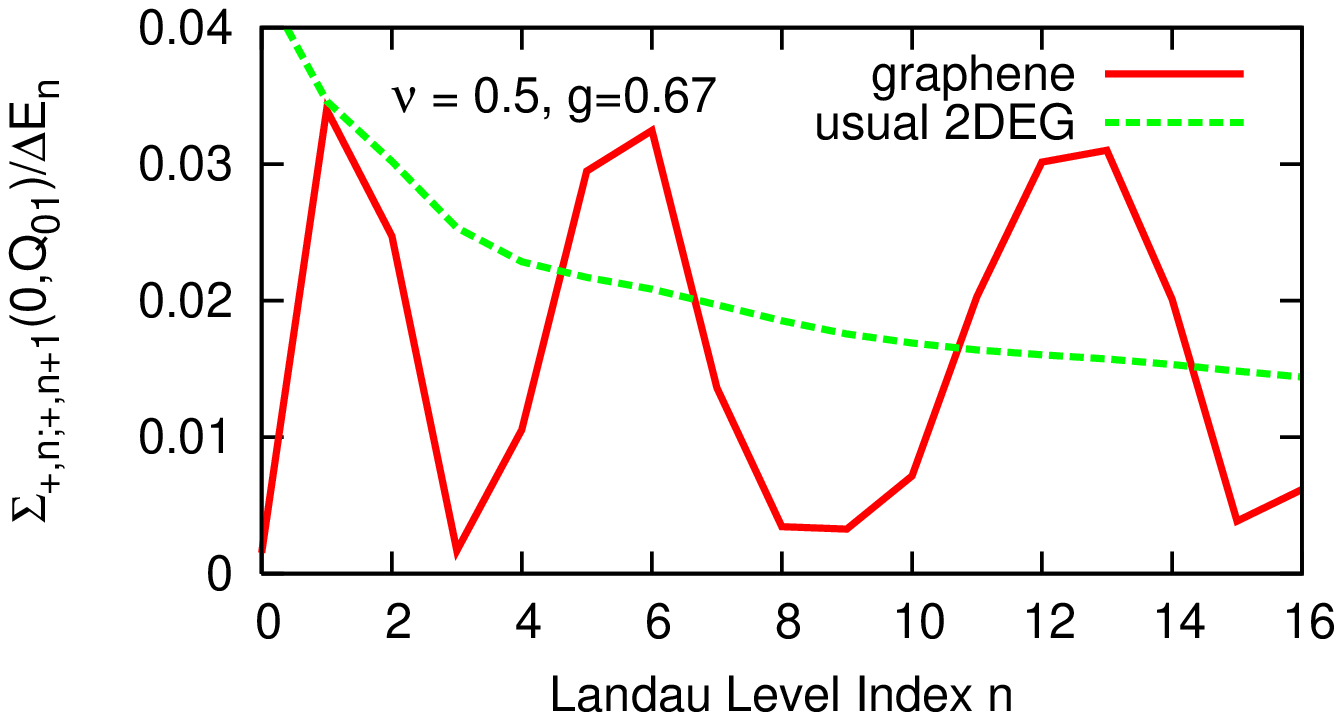}
\end{minipage}
\begin{minipage}{9cm}
\includegraphics[width=9cm]{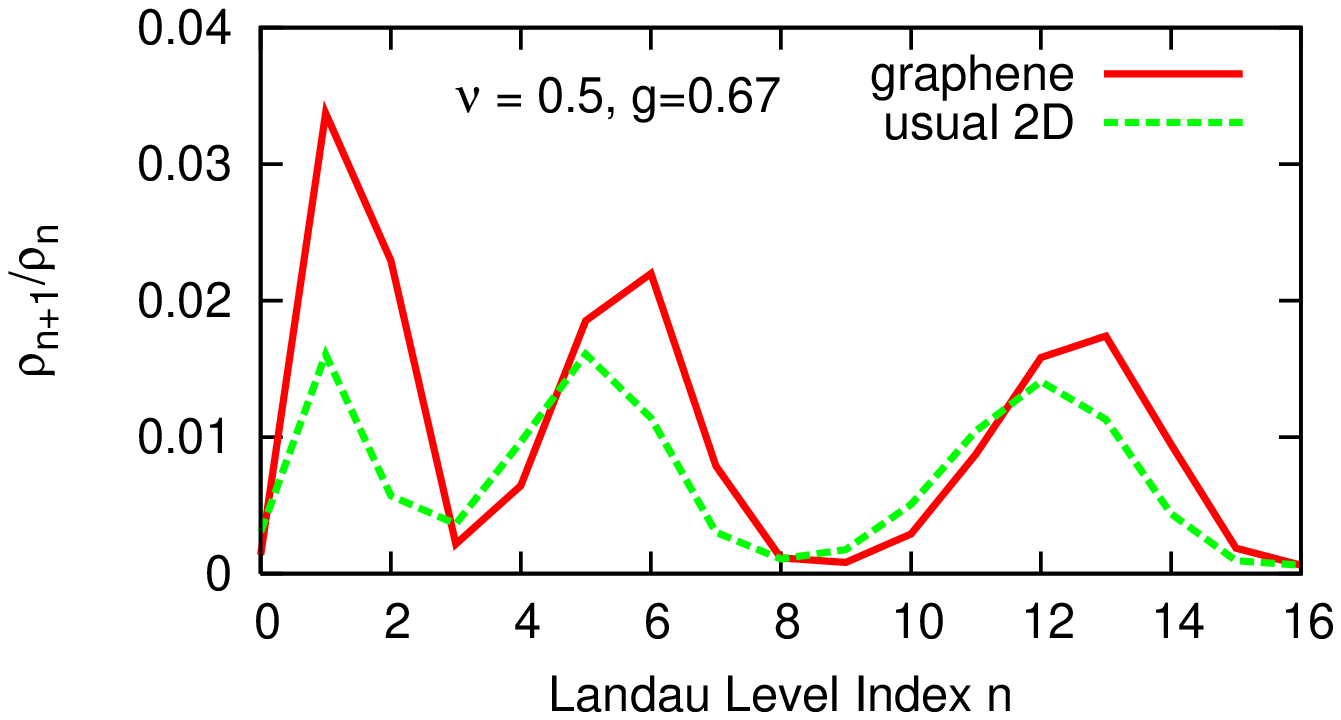}
\end{minipage}
\end{minipage}
\caption{(Color Online) Left: $\Sigma_{+,n;+,n+1}/\Delta E_n$ as a 
function of $n$ for partial 
filling factor $\nu=0.5$ in graphene (solid red) and a conventional 2D 
system (dotted green). In graphene, the ratio is small for all $n$ and 
shows that SLLA is applicable even 
in high Landau levels when $\Delta E_n\rightarrow 0$. In conventional 
2D systems, the ratio decays monotonically as $1/\sqrt{n+1}$ for large 
$n$. Right: Corresponding ratio of occupation numbers $\rho_{n+1}/\rho_n$ 
for graphene 
(solid red) and conventional 2D system (dotted green) provides further 
support for the validity of SLLA in high Landau levels.}
\label{fig:dia_rho}
\end{figure}

Results in Figs.~\ref{fig:mixing} and~\ref{fig:gamma} suggest that the 
effect of Landau-level mixing is not dominant in higher Landau levels, 
even though the energy gap between 
adjacent Landau levels becomes smaller. To understand this unexpected 
result, we recall that the inter-Landau level transitions from $n\rightarrow 
n+1$ are determined by the off-diagonal self-energy matrix elements 
and the gap between adjacent Landau levels, 
$\Sigma_{\sigma n,\sigma n+1}/\Delta E_n$. It follows from 
Eqs.(\ref{eq:sigma},\ref{eq:hartree},\ref{eq:fock}) that for large $n$ 
\begin{align}
\label{eq:asymptotic}
\Sigma_{\sigma n,\sigma n+1}\sim \frac{\rho^{\sigma,\sigma}_{n,n+1}}{(n+1)}
+\frac{a\rho^{\sigma,\sigma}_{n,n}+b\rho^{\sigma,\sigma}_{n+1,n+1}}
{\sqrt{n+1}}
\end{align}
because $\mathcal{F}_{n,n+1}\sim 1/\sqrt{n+1}$. We find that this 
asymptotic behavior is reproduced by our results. We quantify the 
Landau-level mixing by the ratio of relative occupations of the two 
levels in question, $\rho_{n+1}/\rho_{n}$ where 
$\rho_m=\sum_{\sigma}\rho^{\sigma,\sigma}_{m,m}(0)$. Left panel in 
Fig.~\ref{fig:dia_rho} shows the ratio $\Sigma_{n,n+1}/\Delta E_n$ as a 
function of Landau level index $n$ for graphene (solid red) and the 
conventional 2D system (dotted green) at partial filling factor $\nu=0.5$. 
We see that the ratio $\Sigma_{\sigma n,\sigma n+1}(0,\bQ_{01})/\Delta E_n$, 
for typical off-diagonal self-energy matrix element in graphene, is 
smaller than 4\%. In contrast to this, the ratio and the self-energy 
for a conventional 2D system decreases monotonically, since 
$\Delta E_n=\hbar\omega_c$ is independent of $n$, and is well-described by 
a $1/\sqrt{n+1}$ dependence at large $n$. We recall that this ratio for 
a conventional 2D system depends on the magnetic field $B$ or the 
magnetic length $l_B$. Our results are for $g=l_B/a_B=0.67$ or 
$l_B\sim 35$ \AA. (This $g=V_c/\Delta E_n$ for a conventional 2D system is 
equal to the $g=\alpha_G$ in graphene with $\epsilon=3.3$ as the dielectric 
constant.)  The right panel in Fig.~\ref{fig:dia_rho} shows the 
corresponding relative occupations for 
graphene (solid red) and the conventional 2D system (dotted green). 
The fact that this ratio, in the presence of inter-Landau level mixing, is 
small ($\rho_{n+1}/\rho_n\leq 4\%$) provides complementary support for the 
validity of SLLA in graphene.

\section{Discussion}
\label{sec:disc}

In this paper, we have investigated the effects of inter-Landau level 
transitions on Wigner crystal mean-field states in graphene obtained using 
single-Landau-level approximation.~\cite{Zhang07} Our results show  
that the Landau-level mixing does not qualitatively change the phase diagram 
of the system, although it shifts upwards the critical values of filling 
factor $\nu$ at which transitions from one lattice structure to another occur. 
We quantify the Landau-level mixing in terms of off-diagonal self-energy 
and relative occupation numbers, and show that it remains small as a 
function of the Landau level index $n$. 
Thus we conclude that SLLA provides a reliable description of Wigner 
crystal ground states in graphene. 

We emphasize that our results for graphene are independent of the magnetic 
field $B$. For conventional 2D systems, the Landau-level mixing depends 
on the magnetic field and can be important at weak fields $B\leq B_c$ when the 
magnetic length becomes larger than the Bohr radius of the massive carriers, 
$l_B\geq a_B$ for $B\leq B_c$. The absence of a corresponding critical 
field $B_c$ in graphene is due to the massless nature of the carriers. Our 
conclusions do not depend, qualitatively, on the range of the interaction 
$V(\bq)$ because the large-$q$ scattering is strongly suppressed by the 
form factors $\mathcal{F}(\bq)$ that decay exponentially with $q$. 
In this paper, we have ignored transitions to next-higher Landau 
levels~\cite{MacDonald84} $n\rightarrow n+k$, 
because the amplitude for them vanishes 
rapidly: $\Sigma_{n,n+k}/\Delta E_{nk}\sim \sqrt{(n+1)!/(n+k)!k^2}\rightarrow 
0$ as $n\rightarrow\infty$ for $k\geq 2$. This estimate follows from an 
analysis similar to that for Eq.(\ref{eq:asymptotic}) and the observation 
that, in graphene, $\Delta E_{nk}=E_{n+k}-E_n\propto k/\sqrt{n}$ for large 
$n\gg k$. Therefore, it is sufficient to consider the Landau-level mixing 
only between adjacent levels.

Since carriers in graphene are on the surface, in contrast to those in the 
conventional 2D system, it is an ideal candidate for {\it direct 
observation of the local carrier density structure.}~\cite{yacobi} Our results 
provide further support for the existence of triangular Wigner lattice as 
the ground state at small $\nu$ and anisotropic ground states in high 
Landau levels for $\nu\rightarrow 1/2$.~\cite{Zhang07,Wang07} A direct 
observation of carrier 
density in graphene in the quantum Hall regime will verify (or falsify) 
our conclusions. 


\bibliographystyle{apsrev}
\bibliography{ref.graphene_ll_mixing}

\end{document}